\documentstyle[epsf,rotate]{mn}

\newcommand{\be}{\begin{equation}}
\newcommand{\ee}{\end{equation}}

\def\spose#1{\hbox to 0pt{#1\hss}} 
\def\lta{\mathrel{\spose{\lower 3pt\hbox{$\sim$}}\raise 2.0pt\hbox{$<$}}}
\def\gta{\mathrel{\spose{\lower 3pt\hbox{$\sim$}}\raise 2.0pt\hbox{$>$}}}

\begin{document}

\title[Globular Cluster Microlensing]{Globular Clusters as Microlensing Targets}
\author[G. Gyuk and G. Holder]{Geza Gyuk$^1$ and Gilbert P. Holder$^{2}$ \\ 
\\
$^1$S.I.S.S.A., via Beirut 2--4, 34014 Trieste, Italy \\
$^2$Department of Astronomy \& Astrophysics, The University of Chicago, 
Chicago, IL~~60637}
\date{Received ***}
\maketitle

\begin{abstract}
We investigate the possibility of using globular clusters as targets for
microlensing searches. Such searches will be challenging and require more
powerful telescopes than now employed, but are feasible in the 0
future. Although expected event rates are low, we show that the wide
variety of lines of sight to globular clusters greatly enhances the
ability to distinguish between halo models using microlensing observations
as compared to LMC/SMC observations alone. 
\end{abstract}


\section{Introduction}
The over 15 microlensing events observed towards the LMC represent a
triumph of experimental perseverance in the face of seemingly
insurmountable odds \cite{events15}.  However, microlensing has perhaps
opened up more questions than it has resolved. What are the lenses? Where
are they? What fraction of the halo do they constitute? Are they really
part of the extended halo? Even if we could unequivocally state that the
answer to the last question is ``yes'', there are still a number of
unanswered questions.  Present observations can tell us very little about
one of the most important issue: the structure of the baryonic halo.  What
is the shape of the dark baryonic halo?  What is the core radius? Is the
halo really a 1/$r^2$ halo?  Indeed, rather than being elucidated, these
issues of structure cloud the interpretation of the events observed
\cite{ggt2}. Studies show that this is inevitable even with a very large
number of events \cite{cloudy}. The difficulty in obtaining information on
the halo structure stems from two sources. First, the basic microlensing
degeneracies prohibit obtaining direct distance information along the line
of sight. Second, even if such information could be obtained, inverting
one dimensional data to obtain a three dimensional distribution is risky
and prone to error.

Two strategies can be employed to address these problems: using parallax
measurements to break the basic degeneracy \cite{parallax} and using
multiple lines of sight to extract structure information directly. A
parallax satellite (or better a pair) would allow us to determine the
distances to individual lenses and hence have information about the
distribution of mass along the line of sight to the LMC.  Given the
expense of satellites, and the difficulties in translating this to three
dimensional information, it is reasonable to more fully explore the second
possibility. This paper will look at the benefits of multiple
lines of sight. They allow a more direct mapping of the variation of density
with position and thus a determination of such quantities as halo
flattening. Ideally the two methods would complement each other allowing a
full mapping of the halo with greatly lessened theoretical uncertainty.

A variety of possibilities come to mind when alternate lines of sight are
contemplated: 
\begin{enumerate}
\item	Andromeda (M31)
\item	dSph
\item	Globular clusters
\item	bulge
\item	spiral arms
\end{enumerate}
Of these, the first provides only a single line of sight through {\em our}
halo and of course will be complicated (form our point of view) by self
lensing. The second possibility is more attractive. Dwarf spheroidals have
a large number of stars in an environment where crowding is not expected
to be a problem. Unfortunately they are both distant and few. Further,
some are not even as luminous as large globular clusters. Discarding the
most problematic, only a handful remain as likely candidates, reducing
their value as probes of varied lines of sight.  Finally, dwarf
spheroidals are not well-understood. Their observed velocity dispersions
seem to either require a substantial dark matter halo for each dSph or
else indicate that they are extended structures formed from the tidal
disruption of a larger system (see e.g. \cite{dSph}). In either case,
self-lensing may be an important consideration.

The last two directions are expected (and observed in the case of the
bulge) to be dominated by lensing from non-halo populations
\cite{MACHObulge}. Information about halo properties may be difficult to
disentangle from the tail of the disk/bulge lensing
distribution. Globulars are well studied systems, do not have halos of
their own \cite{noglobhalos}, and are small enough so that self lensing is
negligible. Further, they are abundant, relatively close (but not too
close) and contain a respectable number of stars.

The rest of this paper is organized in four sections. The first discusses the
selection of appropriate clusters and touches briefly on the expected
number of stars observable. The next two sections set up characteristic galaxy
models we will explore, discuss how effective cluster microlensing is in
distinguishing between them and compare this to observations of the
SMC. Finally, we conclude with a summary of the expected benefits of
cluster microlensing and what it will take to make it happen.

\section{Selection of Globular Clusters}

Observing globular clusters for microlensing will be difficult. The number
of stars is limited both by the low masses of clusters and the highly
crowded conditions. We discuss these challenges in more detail later, but
for now it should be noted that the ability to observe lensing events in a
globular cluster is only half the story.  The other half is whether such
observations are worth the effort: will the information from observations
along multiple lines of sight yield enough additional information as to
the structure of the halo? Nevertheless, it is clear that not all globular
clusters can be used and that the best candidates must be selected. 

We examined the globular cluster catalog made available by William
E. Harris (McMaster University) at
``http://www.physics.mcmaster.ca/Globular.html''. To be useful for a
microlensing search, a globular cluster must satisfy a number of
criteria. First, it must have a reasonable number of potentially
observable stars. Second, the optical depth must be relatively large, and
third the microlensing rate due to halo objects must not be swamped by
events due to disk and bulge lenses. We examine each of these constraints
in turn.

Ideally, we would have simulated a microlensing search to each of the
$\approx$ 150 known globular clusters, creating images based on realistic
cluster models (core radius, tidal radius, luminosity function, etc.) and
observational parameters (seeing, aperture etc). These simulated images
could then be analysed with a photometric pipeline to determine the
effective number of stars observable for each cluster. Such a procedure
would be very complex and time consuming. Furthermore it would be strongly
dependent on the telescopic parameters and the observing strategies
used. We adopted a very different procedure. Clusters were cut on distance
modulus and total luminosity. Estimates based on standard globular
cluster luminosity functions indicate that a $M_V = -7.5$ cluster should
have $\approx$ 200,000 stars above $M_V=10$. We assume such stars can be
observed out to a distance modulus, $m_D$, of 17.5. We discuss the
challenges of observing stars 200 times fainter than present microlensing
searches in the our final section. We thus consider only clusters with
both distance modulus, $m_D<$17.5 and $M_V<-7.5$. Many stars will be
unobservable due to crowding. On the other hand, the typical cluster we
select will be both brighter and closer, increasing the number of stars
above the limiting magnitude. As a reasonable compromise we adopt 100,000
stars for each cluster. We discuss how our results change as this number
is increased or decreased.

We want our candidate clusters to be worth observing: Clusters with a very
low optical depth in {\em all} models will give us little leverage for
distinguishing between models. Since we do not know a priori what models
might be interesting we used a halo independent measure, the heliocentric
distance. Clusters close to the observer will have narrow, short
microlensing tubes and hence low optical depths. Conversely, distant
clusters are much more likely to have high optical depths. Further, local
clusters are much less useful as a probe of the global structure of the
halo even if they have a high optical depth. We thus adopted a cut on the
distance to the globular clusters, $R_s>8\, {\rm kpc}$. Clusters closer than this
had negligible lensing.

Finally, we wished to concentrate on the halo. Thus clusters for which
non-halo lensing would dominate had to be removed. One could calculate the
expected ratio of disk/bulge lensing versus halo lensing, but this
procedure is fraught with the uncertainties of disk and bulge models. We
took a simpler approach, eliminating those clusters lying within 10$^o$
of the galactic plane.

Twenty clusters survived these cuts with an average luminosity $M_V
\approx -8.5$ and distance modulus $m_D\approx$ 16. Optical depths for a
standard halo model normalized to $\tau_{\rm LMC}=2.5\times 10^{-7}$
ranged from 0.3 to 4.6$\times 10^{-7}$ . The clusters are listed in Table
1.
\begin{table}[hb]
\caption{Globular clusters that passed distance, brightness and
latitude cuts. Distances from the Sun are in kpc. Visual luminosity
(M$_V$) and distance modulus (m$_D$) are in magnitudes. The optical depth
is for a standard halo and is in unit of $10^{-8}$.}
\begin{tabular}{|l||r r r|r r r|}
Cluster & D$_\odot$ & l & b & M$_V$ & m$_D$ & $\tau$ \\
\hline 
 NGC 1261 & 15.2 & 270.5 & -52.10 & -7.68 & 15.97 &  6.4 \\
 NGC 1851 & 11.7 & 244.5 & -35.00 & -8.26 & 15.40 &  3.6 \\
 NGC 1904 & 12.2 & 227.2 & -29.40 & -7.73 & 15.46 &  3.4 \\
 NGC 2808 &  8.9 & 282.2 & -11.30 & -9.26 & 15.46 &  3.2 \\
 NGC 5024 & 18.1 & 333.0 &  79.80 & -8.70 & 16.31 &  9.0 \\
 NGC 5272 &  9.7 &  42.2 &  78.70 & -8.77 & 14.96 &  3.6 \\
 NGC 5286 & 11.3 & 311.6 &  10.60 & -8.67 & 16.01 &  7.4 \\
 NGC 5634 & 24.6 & 342.2 &  49.30 & -7.64 & 17.11 & 21.4 \\
 NGC 5986 & 10.0 & 337.0 &  13.30 & -8.31 & 15.83 &  8.7 \\
 NGC 6093 &  8.4 & 352.7 &  19.50 & -7.85 & 15.18 &  6.2 \\
 NGC 6205 & 28.3 &  73.6 &  40.30 & -7.90 & 17.29 & 15.8 \\
 NGC 6325 &  8.1 &   5.5 &  10.70 & -7.94 & 15.66 &  6.2 \\
 NGC 6342 & 13.8 &   6.7 &  10.20 & -8.35 & 16.60 & 21.0 \\
 NGC 6388 &  8.4 &  21.3 &  14.80 & -8.89 & 16.48 &  5.9 \\
 NGC 6569 & 12.5 & 342.1 & -16.40 & -7.56 & 15.83 & 14.3 \\
 NGC 6712 & 25.4 &   5.6 & -14.10 & -9.89 & 17.49 & 46.0 \\
 NGC 6717 &  8.2 &   0.1 & -17.30 & -7.67 & 14.68 &  6.1 \\
 NGC 6838 & 17.7 &  20.3 & -25.70 & -8.21 & 16.73 & 21.5 \\
 NGC 7006 & 10.0 &  65.0 & -27.30 & -9.07 & 15.27 &  4.5 \\
 NGC 7078 & 11.1 &  53.4 & -35.80 & -8.90 & 15.37 &  5.9
\end{tabular}
\end{table}

\section{Model discrimination}
We know little about the detailed structure of dark halos.  Rotation
curves probe the radial profiles of galaxies, but it is difficult to
determine the relative importance of the dark and luminous
material. Further, rotation curves can only provide information in one
direction.  Flaring of HI gas layers in disks would seem to be a good
probe of the shape of the potential when used in conjunction with the
rotation curves \cite{olling}, but seem to give anomalous results that are
difficult to reconcile with stability concerns or other tracers of the
potentials.  Polar ring galaxies provide a unique opportunity to probe the
shapes of dark matter halos in that there are orthogonal probes of the
potential \cite{Sackett}, but the details of polar rings are uncertain and
these galaxies are almost certainly deeply disturbed systems.  Weak
lensing has not yet borne fruit while simulations are not clear and could
be lacking resolution and/or physics.

Expectations for a 
baryonic halo are even more uncertain. A very rough idea of the 
uncertainties involved can be gained by adopting the following
parametrized model for the halo:
\be
 \rho = \rho_0 \frac{a^\beta + r_0^\beta}{a^\beta + R^\beta + (z/q)^\beta}.
\ee
Such a model has 4 unknowns corresponding to the mass, flattening, core
radius and radial fall-off of the halo. Clearly the parameter space
represented by such models is very large. Since the amount of data from
cluster lensing will be small measuring precise values for the parameters
is not expected. Instead, we will concern ourselves with how well cluster
lensing can discriminate between classes of models. We pick representative
models illustrating various portions of the parameter space.

\begin{table}
\caption{Characteristic models used. The core radii are given in kpc. }
\begin{tabular}{|c | r c c| l}
Model & Core & Index & Axis Ratio \\
\hline
A &  5.0 & 2.0 & 1.0 & Standard halo model\\
B & 20.0 & 2.0 & 1.0 & large core model\\
C &  5.0 & 2.0 & 0.6 & flattened model \\
D &  1.0 & 3.0 & 1.0 & non-halo model\\
E &   -  &  -  &  -  & debris model
\end{tabular}
\end{table}

Our models are summarized in Table 2. The first model is the standard halo
model of Griest (1991), with an isotropic velocity dispersion 
$\sigma\sim156\, {\rm km/s}$.
The second is a similar model, but with a much
larger core radius. Third is a flattened halo as suggested by polar ring
and HI flaring studies, with the velocity dispersion reduced in one
dimension according to the tensor virial theorem. 
Fourth is a more steeply declining halo
based on the stellar halo, with a velocity dispersion given by 
$\sigma\sim127 \, {\rm km/s}$. 
This model is supplemented by a heavy thick disk
which does about 30\% of the lensing. As the lenses are not in the dark
halo proper, we call this the non-halo model. Finally, fifth is a model with no
MACHOs in the halo at all. The LMC lensing is provided by a tidal tail or
a disrupted dwarf galaxy in the line of sight as suggested by Zhao (1997). All models are constructed
to have $\tau_{LMC}=2.0 \times 10^{-7}$ and $\Gamma_{LMC} =4.0\times
10^{-14}{\rm s^{-1}}$. Microlensing observations towards the LMC alone will be
incapable of distinguishing between these models since the distribution of
lens masses is unknown.

To determine how well cluster lensing can distinguish between models, we
proceed in the following manner. First, we pick a ``true'' model S from
the set described above. After computing the microlensing rate towards
each target under this model we generate a realization of S. That is to
say we pick an observed number of events for each target based on the rate
and ten years of observation, using Poisson statistics. For each
realization we compute the likelihood for each of our models. The most
likely model will not necessarily be the ``true'' model. After generating
many (100,000) realizations we can build up the probability P(T) that the
model S will be identified by a microlensing experiment as the model
T. Repeating but with a different ``true'' model we build the full matrix
P(S,T) which gives the chance that the real galaxy S will appear to the
observers as T. If P(S,T) is diagonal then we know that the observations
are doing a good jobs of distinguishing models. If on the other hand, it
is far from diagonal, then observations are doing a poor job and
misidentifications are frequent: the data cannot even distinguish between,
for example, a flattened halo and a spherical halo.

We present results from a number of scenarios: LMC/SMC lensing alone, LMC
and globular cluster lensing, and LMC/SMC/cluster/dSph lensing. 
For the dSph lensing, we looked at lensing information that could be
gained from looking at Fornax, Sculptor, Sextans and Ursa Minor. The other
dSph's were either too far, too faint, or some combination of the two
to be useful.
We take
the effective number of stars in the SMC to be $2.0 \times 10^6$. We also
show how these the effectiveness of globular cluster lensing varies with
number of observable stars.

\begin{table}
\caption{Discrimination matrix for five scenarios. See text and Table 2 for
explanation of the models used.}
\begin{tabular}{|r | r r r r r|}
\multicolumn{6}{|l|}{SMC only (2.0e6 stars)} \\
\hline
 &A~~&B~~&C~~&D~~&E~~ \\
\hline
A&0.29 & 0.23 & 0.20 & 0.29 & 0.00 \\ 
B&0.22 & 0.31 & 0.33 & 0.14 & 0.00 \\
C&0.10 & 0.19 & 0.64 & 0.06 & 0.00 \\
D&0.19 & 0.06 & 0.07 & 0.69 & 0.00 \\
E&0.00 & 0.00 & 0.00 & 0.00 & 1.00 \\
& & & & & \\
\multicolumn{6}{|l|}{G.Clusters (2.0e6 stars)} \\
\hline
 &A~~&B~~&C~~&D~~&E~~ \\
\hline
A&0.63 & 0.04 & 0.33 & 0.00 & 0.00 \\
B&0.03 & 0.97 & 0.00 & 0.00 & 0.00 \\  
C&0.35 & 0.01 & 0.64 & 0.00 & 0.00 \\  
D&0.00 & 0.00 & 0.00 & 1.00 & 0.00 \\  
E&0.00 & 0.00 & 0.00 & 0.00 & 1.00 \\  
& & & & & \\
\multicolumn{6}{|l|}{SMC, G. Clusters and dSph} \\
\hline
 &A~~&B~~&C~~&D~~&E~~ \\
\hline
A&0.78 & 0.03 & 0.19 & 0.00 & 0.00 \\  
B&0.02 & 0.98 & 0.00 & 0.00 & 0.00 \\  
C&0.18 & 0.01 & 0.82 & 0.00 & 0.00 \\  
D&0.00 & 0.00 & 0.00 & 1.00 & 0.00 \\  
E&0.00 & 0.00 & 0.00 & 0.00 & 1.00 \\
& & & & & \\
\multicolumn{6}{|l|}{G. Clusters (1.0e6 stars)} \\
\hline
 &A~~&B~~&C~~&D~~&E~~ \\
\hline
A&0.52 & 0.11 & 0.37 & 0.00 & 0.00 \\
B&0.07 & 0.92 & 0.01 & 0.00 & 0.00 \\
C&0.35 & 0.06 & 0.59 & 0.00 & 0.00 \\
D&0.00 & 0.00 & 0.00 & 1.00 & 0.00 \\
E&0.00 & 0.00 & 0.00 & 0.00 & 1.00 \\
& & & & & \\
\multicolumn{6}{|l|}{G. Clusters (4.0e6 stars)} \\
\hline
 &A~~&B~~&C~~&D~~&E~~ \\
\hline
A&0.72 & 0.01 & 0.28 & 0.00 & 0.00 \\
B&0.00 & 1.00 & 0.00 & 0.00 & 0.00 \\
C&0.29 & 0.00 & 0.71 & 0.00 & 0.00 \\
D&0.00 & 0.00 & 0.00 & 1.00 & 0.00 \\
E&0.00 & 0.00 & 0.00 & 0.00 & 1.00 
\end{tabular}
\end{table}

\section{Results}

It has been suggested that SMC lensing can break the degeneracy between
flattened and spherical halos (eg.  Sackett \& Gould 1993). Unfortunately,
for realistic event rates the time required is large. As can be seen in
Table 3, for a ten year experiment LMC/SMC observations can correctly
select the flattened model only $\approx$60\% of the time. Examination of
the likelihoods directly shows that even when the correct model is
selected, the margin of probability is small with spherical models and
large core radii models only slightly less preferred. When the ``true''
model has a large core SMC lensing performs abysmally. Indeed, most of the
time another model is preferred. 
As is true of the LMC, the line of
sight towards the SMC is most sensitive at large galactocentric distances,
where the effect of a larger core radius is small. As expected,  SMC lensing
does better for non-halo models. Here the direction of the SMC line of
sight closer to the Galactic center allows it to penetrate into the dense
inner regions. However, there is still a reasonable chance (30\%) that an
unexpectedly high amount of SMC lensing mimics one of the other
models. Further, examination of the likelihoods again shows that the
probabilities are not sharply peaked on the non-halo models. Thus our
confidence in this determination will not be high. The debris models are
always correctly identified. LMC/SMC lensing in general does not fare well
in distinguishing between models because it only probes two lines of
sight. This is particularly worrisome if self lensing is important as has
been suggested for the SMC \cite{EROSSMC}. Even a small amount of unknown
self lensing is capable of throwing the model determinations into grave
doubt. Furthermore, if the detailed structure of the baryonic halo is
lumpy, as suggested in some models, the determinations become even more
uncertain. Microlensing towards the SMC appears to be a poor tool for
elucidating halo structure.

Globular cluster microlensing can avoid some of these problems. First of
all, quick calculations show that $\tau_{self} \ll 10^{-8}$ for all
globular clusters. Thus, self lensing will be negligible. Second, since
many more lines of sight are involved, lumpiness in the detailed
distribution of the mass will be largely smoothed over. The figures for
LMC/globular cluster lensing shown in Table 3 should therefore  be more robust
than the corresponding figures for LMC/SMC lensing. 

\begin{figure}
\epsfysize=10.0cm
\centerline{
\rotate[r]{\epsfbox{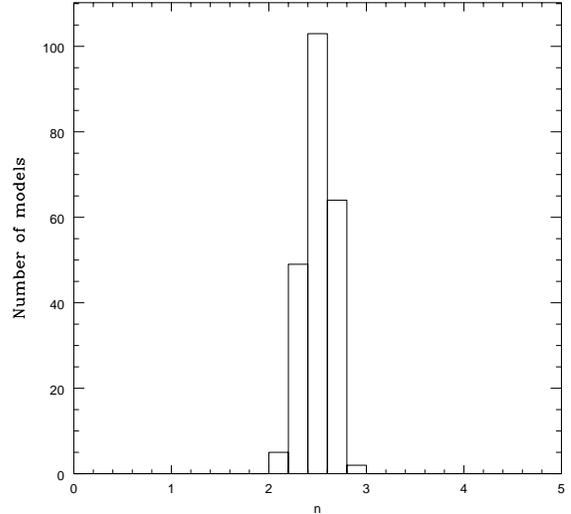}}}
\caption{Distribution of best-fit power law indices for a true model of
n=2.5 and LMC/SMC/cluster/dSph lensing. Clusters are assumed to have
100,000 stars monitorable.}
\end{figure}

A quick examination shows that globular cluster lensing is much better at
distinguishing between models with different core radii. This is quite
understandable: globular clusters probe precisely the region where
differences in the core are most apparent. Similarly, the selection of the
the non-halo models certain. Models with a large power-law index predict
much higher lensing towards the globular clusters making the distinction
very easy and certain. LMC/globular cluster observations do not fare 
better than LMC/SMC observations at distinguishing flattened models. This
is because only a few clusters are at high galactic latitude. As before,
the debris models are firmly ruled out. We also examined a scenario where
the SMC, globular clusters and four dSph galaxies were all
observed. For these purposes we assumed that it was possible to monitor
500,000 stars in each dSph. Results are shown in the third part of Table
3. The improvement in the determination of the flattening is considerable
due to the relatively high latitudes of the dSph's. 

Since the core radius and the power-law index are so easily distinguished
by the full set of targets, we investigated the question of 
how well quantitative
information could be obtained. We picked a standard model with
n=2.5. Realizations of this model were then compared to models where the
optical depth towards the LMC was held fixed but the power law index was
allowed to float. Figure 1 shows the distribution of power law indices
inferred. The peak is clearly around the ``true'' value with a small
($\approx 0.2$) variance. The concentration of the globular clusters
towards the Galactic Center allows very accurate determination of this
parameter. A similar test was performed for the core radius, this time
with a core of 12 kpc. Figure 2 shows the distribution of inferred
values. Note that the distribution is not quite as tight ($\sigma \approx
5$ kpc), but still allows fairly accurate determination of the core radius.

\begin{figure}
\epsfysize=10.0cm
\centerline{
\rotate[r]{\epsfbox{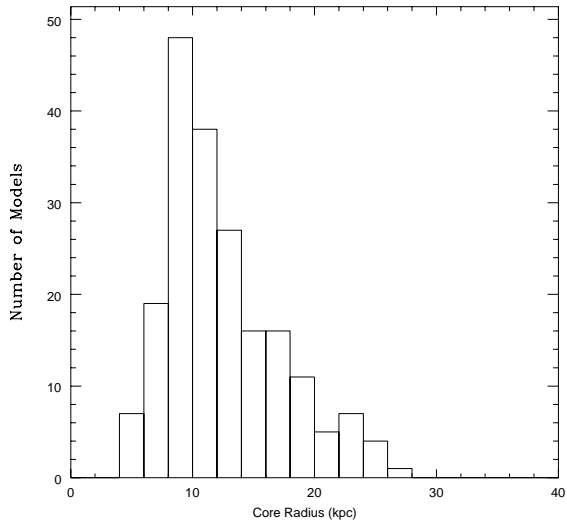}}}
\caption{Distribution of best-fit core radii for a true model of
a=12 kpc and LMC/SMC/cluster/dSph lensing. Clusters are assumed to have
100,000 stars monitorable.}
\end{figure}

Finally, we examined how much our results depended on the number of stars
observable per cluster. Table 4 shows the results of LMC/globular cluster
lensing for 50,000 and 200,000 stars per cluster. We see that the results
are not strongly sensitive to increases in the number of stars observed,
although marginal improvement is obtained. When the number of stars is
decreased, however, the returns on cluster lensing are lower. Nevertheless, 
clusters can still discriminate between models better than LMC/SMC lensing
especially when the possible problems with LMC/SMC lensing are
considered. Globular cluster microlensing is a 
worthwhile investment, even at lower
numbers of effective stars.

\section{Conclusions}

Globular cluster microlensing searches are much  
better for determining important halo parameters
than LMC/SMC measurements alone. Although few events are involved, the
diverse lines of sight allow far greater leverage on halo structure
parameters. In particular, the power-law index and the core radius are
readily accessible to cluster data. 

The importance of diverse lines of sight has already been demonstrated:
reports of a microlensing event seen towards the SMC by both the MACHO and
EROS collaborations \cite{EROSSMC,MACHOSMC} suggest that the debris models
for LMC lensing are not viable. Caution must be taken however, as the self
lensing due to the SMC could cloud this interpretation. The small size and
low self-lensing of the globular clusters is a definite advantage in this
respect.

The faintness of the typical globular cluster stars ($M_V\approx 26$), and
the highly crowded conditions ($>$100 stars/arcsec$^2$ at the core) both
make microlensing observations difficult. This suggests that the best bet
for cluster microlensing studies would be a large ($\approx 8m$) telescope
with excellent seeing ($< 0.5''$), possibly achieved with adaptive
optics. Telescopes of this type, such as the SUBARU, are just now coming
on-line. Indeed, Yanagisawa \& Muraki (1997), have done a preliminary
study and conclude that the SUBARU telescope should be able to monitor
about 200,000 stars each in Pal 8 and NGC6453 with 10 second
exposures. Longer exposures and stellar templates provided by HST may be
able to push this number up considerably. Even if such large telescopes
and high resolution are not available, application of the pixel lensing
technique may allow monitoring of sufficient stars.  Rhoads \& Malhotra
(1997) calculate that pixel lensing can allow 200,000 star equivalents to
be monitored in M15 and an average over all globular clusters of about
60,000. We note that our clusters are considerably more luminous than the
average cluster. Finally, it is important to realize that such experiments
have {\em already} been performed! Mighell et al. (1992) reported two
nights of observations toward Omega Centauri with the Anglo-Australian
telescope. Although no events were detected, 50,000 stars in a 12.5' field
were monitored with 10 sec exposures. In an ongoing effort to search for
variable stars, the OGLE collaboration has been examining portions of
Omega Centauri and 47 Tuc \cite{OGLECen}. With 500 sec exposures in 1.6''
seeing they follow about 30,000 stars in Omega Centauri. In light of these
numbers, the target figure of 100,000 stars per cluster seems reasonable.

The challenges involved in cluster microlensing searches should not be
underestimated: following millions of very faint stars in extremely crowded
fields scattered over a large part of the sky will require dedication,
both of telescope time and its practitioners. The telescopes and
techniques are both at the cutting edge of astronomy. However, optimism is
warranted: 10 years ago microlensing was a pipe dream, now seven collaborations
are actively pursuing it and millions of stars are monitored nightly.
We must be bold in proposing and following up new ideas.

\section*{Acknowledgements}

After this work was completed we received a preprint
(astro-ph/9710231, Rhoads \& Malhotra 1997) which reaches similar
conclusions.

\end{document}